\documentclass[10pt,twocolumn]{IEEEtran}
\usepackage{times}
\usepackage{amsbsy}
\usepackage{latexsym}
\usepackage{amsmath}
\usepackage[ansinew]{inputenc}
\usepackage[dvips ]{graphicx}
\input epsf
\usepackage{epic,eepic}
\usepackage{url}
\usepackage{algorithm}
\usepackage{color}
\usepackage{booktabs}
\usepackage{setspace}
\usepackage{listings}
\usepackage[usenames,dvipsnames,svgnames,table]{xcolor}
\usepackage{cite}
\input epsf
\title{\Huge Detection and Estimation Algorithms in Massive MIMO Systems }
\author{Rodrigo C. de Lamare and Raimundo Sampaio-Neto \\ Centre for Telecommunications Studies (CETUC) \\ Pontifical Catholic University of Rio de Janeiro, Gávea - 22453-900, Rio de Janeiro, Brazil \\ Communications Research Group \\ Department of Electronics,
    University of York, York Y010 5DD, United Kingdom \\
    Email: \protect\url{delamare@cetuc.puc-rio.br,raimundo@cetuc.puc-rio.br}
\thanks{\footnotesize The work of the authors is
funded by the Pontifical Catholic University of Rio de Janeiro and
CNPq. }}

\begin{document}
\maketitle

\begin{abstract}

This book chapter reviews signal detection and parameter estimation
techniques for multiuser multiple-antenna wireless systems with a
very large number of antennas, known as massive multi-input
multi-output (MIMO) systems. We consider both centralized antenna
systems (CAS) and distributed antenna systems (DAS) architectures in
which a large number of antenna elements are employed and focus on
the uplink of a mobile cellular system. In particular, we focus on
receive processing techniques that include signal detection and
parameter estimation problems and discuss the specific needs of
massive MIMO systems. Simulation results illustrate the performance
of detection and estimation algorithms under several scenarios of
interest. Key problems are discussed and future trends in massive
MIMO systems are pointed out.

\end{abstract}

\section{Introduction}

Future wireless networks will have to deal with a substantial
increase of data transmission due to a number of emerging
applications that include machine-to-machine communications and
video streaming \cite{cisco}-\cite{5g}. This very large amount of
data exchange is expected to continue and rise in the next decade or
so, presenting a very significant challenge to designers of
fifth-generation (5G) wireless communications systems \cite{5g}.
Amongst the main problems are how to make the best use of the
available spectrum and how to increase the energy efficiency in the
transmission and reception of each information unit. 5G
communications will have to rely on technologies that can offer a
major increase in transmission capacity as measured in bits/Hz/area
but do not require increased spectrum bandwidth or energy
consumption.

Multiple-antenna or multi-input multi-output (MIMO) wireless
communication devices that employ antenna arrays with a very large
number of antenna elements which are known as massive MIMO systems
have the potential to overcome those challenges and deliver the
required data rates, representing a key enabling technology for 5G
\cite{marzetta_first}-\cite{nam}. Among the devices of massive MIMO
networks are user terminals, tablets, machines and base stations
which could be equipped with a number of antenna elements with
orders of magnitude higher than current devices. Massive MIMO
networks will be structured by the following key elements: antennas,
electronic components, network architectures, protocols and signal
processing. The network architecture, in particular, will evolve
from homogeneous cellular layouts to heterogeneous architectures
that include small cells and the use of coordination between cells
\cite{combes}. Since massive MIMO will be incorporated into mobile
cellular networks in the future, the network architecture will
necessitate special attention on how to manage the interference
created \cite{aggarwal} and measurements campaigns will be of
fundamental importance \cite{shepard}-\cite{gao}. The coordination
of adjacent cells will be necessary due to the current trend towards
aggressive reuse factors for capacity reasons, which inevitably
leads to increased levels of inter-cell interference and signalling.
The need to accommodate multiple users while keeping the
interference at an acceptable level will also require significant
work in scheduling and medium-access protocols.

Another important aspect of massive MIMO networks lies in the signal
processing, which must be significantly advanced for 5G. In
particular, MIMO signal processing will play a crucial role in
dealing with the impairments of the physical medium and in providing
cost-effective tools for processing information. Current
state-of-the-art in MIMO signal processing requires a computational
cost for transmit and receive processing that grows as a cubic or
super-cubic function of the number of antennas, which is clearly not
scalable with a large number of antenna elements. We advocate the
need for simpler solutions for both transmit and receive processing
tasks, which will require significant research effort in the next
years. Novel signal processing strategies will have to be developed
to deal with the problems associated with massive MIMO networks like
computational complexity and its scalability, pilot contamination
effects, RF impairments, coupling effects, delay and calibration
issues.

In this chapter, we focus on signal detection and parameter
estimation aspects of massive MIMO systems. We consider both
centralized antenna systems (CAS) and distributed antenna systems
(DAS) architectures in which a large number of antenna elements are
employed and focus on the uplink of a mobile cellular system. In
particular, we focus on the uplink and receive processing techniques
that include signal detection and parameter estimation problems and
discuss specific needs of massive MIMO systems. We review the
optimal maximum likelihood detector, nonlinear and linear suboptimal
detectors and discuss potential contributions to the area. We also
describe iterative detection and decoding algorithms, which exchange
soft information in the form of log likelihood ratios (LLRs) between
detectors and channel decoders. Another important area of
investigation includes parameter estimation techniques, which deal
with methods to obtain the channel state information, compute the
parameters of the receive filters and the hardware mismatch.
Simulation results illustrate the performance of detection and
estimation algorithms under scenarios of interest. Key problems are
discussed and future trends in massive MIMO systems are pointed out.

This chapter is structured as follows. Section II reviews the signal
models with CAS and DAS architectures and discusses the application
scenarios. Section III is dedicated to detection techniques, whereas
Section IV is devoted to parameter estimation methods. Section V
discusses the results of some simulations and Section VI presents
some open problems and suggestions for further work. The conclusions
of this chapter are given in Section VII.

\section{Signal Models and Application Scenarios}

In this section, we describe signal models for the uplink of
multiuser massive MIMO systems in mobile cellular networks. In
particular, we employ a linear algebra approach to describe the
transmission and how the signals are collected at the base station
or access point. We consider both CAS and DAS
\cite{das_dai,choi_andrews} configurations. In the CAS configuration
a very large array is employed at the rooftop or at the façade of a
building or even at the top of a tower. In the DAS scheme,
distributed radio heads are deployed over a given geographic area
associated with a cell and these radio devices are linked to a base
station equipped with an array through either fibre optics or
dedicated radio links. These models are based on the assumption of a
narrowband signal transmission over flat fading channels which can
be easily generalized to broadband signal transmission with the use
of multi-carrier systems.

\begin{figure}[!htb]
\begin{center}
\def\epsfsize#1#2{1\columnwidth}
\epsfbox{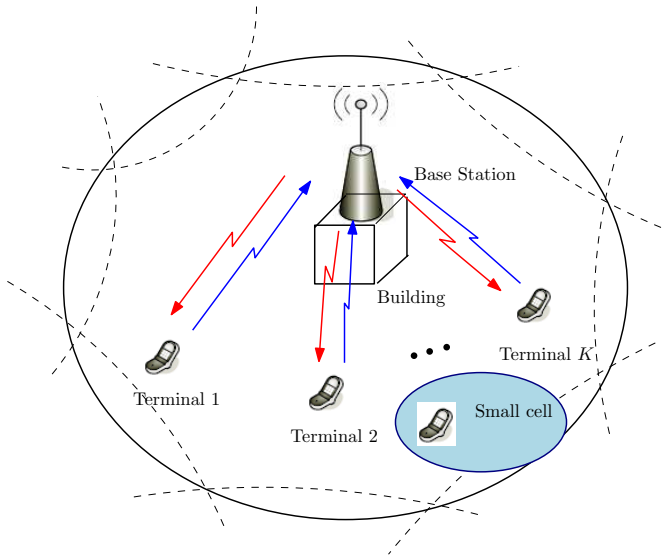} \vspace{-1em} \caption{Mobile cellular network
with a CAS configuration.}\label{fig1}
\end{center}
\end{figure}

The scenario we are interested in this work is that of mobile
cellular networks beyond LTE-A \cite{lte-a} and 5G communications
\cite{5g}, which is illustrated in Fig. \ref{fig1}. In such
networks, massive MIMO will play a key role with the deployment of
hundreds of antenna elements at the base station using CAS or using
DAS over the cell of interest, coordination between cells and a more
modest number of antenna elements at the user terminals. At the base
station, very large antenna arrays could be deployed on the roof or
on the façade of buildings. With further development in the area of
compact antennas and techniques to mitigate mutual coupling effects,
it is likely that the number of antenna elements at the user
terminals (mobile phones, tables and other gadgets) might also be
significantly increased from $1-4$ elements in current terminals to
$10-20$ in future devices. In these networks, it is preferable to
employ time-division-duplexing (TDD) mode to perform uplink channel
estimation and obtain downlink CSI by reciprocity for signal
processing at the transmit side. This operation mode will require
cost-effective calibration algorithms. Another critical requirement
is the uplink channel estimation, which employs non-orthogonal
pilots and, due to the existence of adjacent cells and the coherence
time of the channel, needs to reuse the pilots \cite{jose}. Pilot
contamination occurs when CSI at the base station in one cell is
affected by users from other cells. In particular, the uplink (or
multiple-access channel) will need CSI obtained by uplink channel
estimation, efficient multiuser detection and decoding algorithms.
The downlink (also known as the broadcast channel) will require CSI
obtained by reciprocity for transmit processing and the development
of cost-effective scheduling and precoding algorithms. A key
challenge in the scenario of interest is how to deal with a very
large number of antenna elements and develop cost-effective
algorithms, resulting in excellent performance in terms of the
metrics of interest, namely, bit error rate (BER), sum-rate and
throughput. In what follows, signal models that can describe CAS and
DAS schemes will be detailed.

\subsection{Centralized Antenna System Model}

In this subsection, we consider a multiuser massive MIMO system with
CAS using $N_A$ antenna elements at the receiver, which is located
at a base station of a cellular network installed at the rooftop of
a building or a tower, as illustrated in Fig. \ref{fig1}. Following
this description, we consider a multiuser massive MIMO system with
$K$ users that are equipped with $N_U$ antenna elements and
communicate with a receiver with $N_A$ antenna elements, where $N_A
\geq K N_U$. At each time instant, the $K$ users transmit $N_U$
symbols which are organized into a $N_U \times 1$ vector
${\boldsymbol s}_k [i] = \big[ s_{k,1}[i], ~s_{k,2}[i], ~ \ldots,~
s_{k,N_U}[i] \big]^T$ taken from a modulation constellation $A = \{
a_1,~a_2,~\ldots,~a_N \}$. The data vectors ${\boldsymbol s}_k[i]$
are then transmitted over flat fading channels. The received signal
after demodulation, pulse-matched filtering and sampling is
collected in an $N_A \times 1$ vector ${\boldsymbol r}[i] = \big[
r_1[i], ~r_2[i], ~ \ldots,~ r_{N_A}[i] \big]^T$ with sufficient
statistics for estimation and detection as described by
\begin{equation}
\begin{split}
{\boldsymbol r}[i] & = \sum_{k=1}^{K} \gamma_k {\boldsymbol H}_k
{\boldsymbol s}_k[i] + {\boldsymbol n}[i] \\
& = \sum_{k=1}^{K}{\boldsymbol G}_k {\boldsymbol s}_k[i] +
{\boldsymbol n}[i],
\end{split}
\end{equation}
where the $N_A \times 1$ vector ${\boldsymbol n}[i]$ is a zero mean
complex circular symmetric Gaussian noise with covariance matrix
$E\big[ {\boldsymbol n}[i] {\boldsymbol n}^H[i] \big] = \sigma_n^2
{\boldsymbol I}$. The data vectors ${\boldsymbol s}_k[i]$ have zero
mean and covariance matrices $E\big[ {\boldsymbol s}_k[i]
{\boldsymbol s}_k^H[i] \big] = \sigma_{s_{k}}^2 {\boldsymbol I}$,
where $\sigma_{s_{k}}^2$ is the user $k$ transmit signal power. The
elements $h_{i,j}^k$ of the $N_A \times N_U$ channel matrices
${\boldsymbol H}_k$ are the complex channel gains from the $j$th
transmit antenna of user $k$ to the $i$th receive antenna. For a CAS
architecture, the channel matrices ${\boldsymbol H}_k$ can be
modeled using the Kronecker channel model \cite{kermoal} as detailed
by
\begin{equation}
{\boldsymbol H}_{k} = {\boldsymbol \Theta}_R^{1/2} {\boldsymbol
H}_{k}^{o} {\boldsymbol \Theta}_{T}^{1/2},
\end{equation}
where ${\boldsymbol H}_{k}^{o}$ has complex channel gains obtained
from complex Gaussian random variables with zero mean and unit
variance, ${\boldsymbol \Theta}_R$ and ${\boldsymbol \Theta}_{T}$
denote the receive and transmit correlation matrices, respectively.
The components of correlation matrices ${\boldsymbol \Theta}_R$ and
${\boldsymbol \Theta}_T$ are of the form
\begin{equation}
{\boldsymbol \Theta}_{R/T} = \left(
                             \begin{array}{ccccc}
                               1 & \rho & \rho^4 & \ldots & \rho^{(N_a-1)^2} \\
                               \rho & 1 & \rho & \ldots & \vdots \\
                               \rho^4 & \rho & 1 & \vdots &  \rho^4 \\
                               \vdots & \vdots & \vdots & \vdots & \vdots \\
                               \rho^{(N_a-1)^2} & \ldots & \rho^4 & \rho & 1 \\
                             \end{array}
                           \right)
\end{equation}
where $\rho$ is the correlation index of neighboring antennas and
$N_a$ is the number of antennas of the transmit or receive array.
When $\rho = 0$ we have an uncorrelated scenario and when $\rho = 1$
we have a fully correlated scenario. The channels between the
different users are assumed uncorrelated due to their geographical
location.

The parameters $\gamma_k$ represent the large-scale propagation
effects for user $k$ such as path loss and shadowing which are
represented by
\begin{equation}
\gamma_k = \alpha_k \beta_k,
\end{equation}
where the path loss $\alpha_k$ for each user is computed by
\begin{equation}
\alpha_k = \sqrt{\frac{L_{k}}{d_k^{\tau}}},
\end{equation}
where $L_{k}$ is the power path loss of the link associated with
user $k$, $d_k$ is the relative distance between the user and the
base station, $\tau$ is the path loss exponent chosen between $2$
and $4$ depending on the environment.

The log-normal shadowing $\beta_k$ is given by
\begin{equation}
\beta_k = 10^{\frac{\sigma_k v_k}{10}},
\end{equation}
where $\sigma_k$ is the shadowing spread in dB and $v_k$ corresponds
to a real-valued Gaussian random variable with zero mean and unit
variance. The $N_A \times N_U$ composite channel matrix that
includes both large-scale and small-scale fading effects is denoted
as ${\boldsymbol G}_k$. 

\subsection{Distributed Antenna Systems Model}

\begin{figure}[!htb]
\begin{center}
\def\epsfsize#1#2{1\columnwidth}
\epsfbox{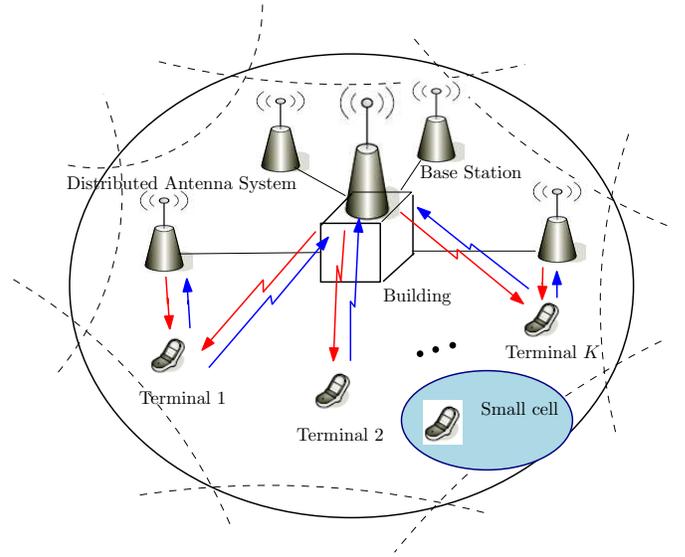} \vspace{-1em} \caption{Mobile cellular network
with a DAS configuration.}\label{fig2}
\end{center}
\end{figure}

In this subsection, we consider a multiuser massive MIMO system with
a DAS configuration using $N_B$ antenna elements at the base station
and $L$ remote radio heads each with $Q$ antenna elements, which are
distributed over the cell and linked to the base station via wired
links, as illustrated in Fig. \ref{fig2}. Following this
description, we consider a multiuser massive MIMO system with $K$
users that are equipped with $N_U$ antenna elements and communicate
with a receiver with a DAS architecture with a total of $N_A=N_B+LQ$
antenna elements, where $N_A \geq K N_U$. In our exposition, when
the number of remote radio heads is set to zero, i.e., $L=0$, the
DAS architecture reduces to the CAS scheme with $N_A = N_B$.

At each time instant, the $K$ users transmit $N_U$ symbols which are
organized into a $N_U \times 1$ vector ${\boldsymbol s}_k [i] =
\big[ s_{k,1}[i], ~s_{k,2}[i], ~ \ldots,~ s_{k,N_U}[i] \big]^T$
taken from a modulation constellation $A = \{ a_1,~a_2,~\ldots,~a_N
\}$. The data vectors ${\boldsymbol s}_k[i]$ are then transmitted
over flat fading channels. The received signal after demodulation,
pulse-matched filtering and sampling is collected in an $N_A \times
1$ vector ${\boldsymbol r}[i] = \big[ r_1[i], ~r_2[i], ~ \ldots,~
r_{N_A}[i] \big]^T$ with sufficient statistics for estimation and
detection as described by
\begin{equation}
\begin{split}
{\boldsymbol r}[i] & = \sum_{k=1}^{K} {\boldsymbol \gamma}_k
{\boldsymbol H}_k {\boldsymbol s}_k[i] + {\boldsymbol n}[i] \\
 & = \sum_{k=1}^{K} {\boldsymbol G}_k {\boldsymbol
s}_k[i] + {\boldsymbol n}[i],
\end{split}
\end{equation}
where the $N_A \times 1$ vector ${\boldsymbol n}[i]$ is a zero mean
complex circular symmetric Gaussian noise with covariance matrix
$E\big[ {\boldsymbol n}[i] {\boldsymbol n}^H[i] \big] = \sigma_n^2
{\boldsymbol I}$. The data vectors ${\boldsymbol s}_k[i]$ have zero
mean and covariance matrices $E\big[ {\boldsymbol s}_k[i]
{\boldsymbol s}_k^H[i] \big] = \sigma_{s_{k}}^2 {\boldsymbol I}$,
where $\sigma_{s_{k}}^2$ is the user $k$ signal power. The elements
$h_{i,j}$ of the $N_A \times N_U$ channel matrices ${\boldsymbol
H}_k$ are the complex channel gains from the $j$th transmit antenna
to the $i$th receive antenna. Unlike the CAS architecture, in a DAS
setting the channels between remote radio heads are less likely to
suffer from correlation due to the fact that they are geographically
separated. However, for the antenna elements located at the base
station and at each remote radio head, the $L+1$ submatrices of
${\boldsymbol H}_k$ can be modeled using the Kronecker channel model
\cite{kermoal} as detailed in the previous subsection. The major
difference between CAS and DAS schemes lies in the large-scale
propagation effects. Specifically, with DAS the links between the
users and the distributed antennas experience in average lower path
loss effects because of the reduced distance between their antennas.
This helps to create better wireless links and coverage of the cell.
Therefore, the large-scale propagation effects are modeled by an
$N_A \times N_A$ diagonal matrix given by
\begin{equation}
{\boldsymbol \gamma}_k = {\rm diag} \left( \underbrace{\gamma_{k,1}
\ldots \gamma_{k,1}}_{N_B}~
\underbrace{\gamma_{k,2} \ldots \gamma_{k,2}}_{Q} ~ \ldots ~\underbrace{\gamma_{k,L+1} \ldots \gamma_{k,L+1}}_{Q}  \right), 
\end{equation}
where the parameters $\gamma_{k,j}$ for $j=1, \ldots, L+1$ denote
the large-scale propagation effects like shadowing and pathloss from
the $k$th user to the $j$th radio head. The parameters
$\gamma_{k,j}$ for user $k$ and distributed antenna $j$ are
described by
\begin{equation}
\gamma_{k,j} = \alpha_{k,j} \beta_{k,j},~~ j=1, \ldots, L+1
\end{equation}
where the path loss $\alpha_{k,j}$ for each user and antenna is
computed by
\begin{equation}
\alpha_{k,j} = \sqrt{\frac{L_{k,j}}{d_{k,j}^{\tau}}},
\end{equation}
where $L_{k,j}$ is the power path loss of the link associated with
user $k$ and the $j$th radio head, $d_{k,j}$ is the relative
distance between the user and the radio head, $\tau$ is the path
loss exponent chosen between $2$ and $4$ depending on the
environment. The log-normal shadowing $\beta_{k,j}$ is given by
\begin{equation}
\beta_{k,j} = 10^{\frac{\sigma_k v_{k,j}}{10}},
\end{equation}
where $\sigma_k$ is the shadowing spread in dB and $v_{k,j}$
corresponds to a real-valued Gaussian random variable with zero mean
and unit variance. The $N_A \times N_U$ composite channel matrix
that includes both large-scale and small-scale fading effects is
denoted as ${\boldsymbol G}_k$.

\section{Detection Techniques}

In this section, we examine signal detection algorithms for massive
MIMO systems. In particular, we review various detection techniques
and also describe iterative detection and decoding schemes that
bring together detection algorithms and error control coding.

\subsection{Detection Algorithms}

In the uplink of the multiuser massive MIMO systems under
consideration, the signals or data streams transmitted by the users
to the receiver overlap and typically result in multiuser
interference at the receiver. This means that the interfering
signals cannot be easily demodulated at the receiver unless there is
a method to separate them. In order to separate the data streams
transmitted by the different users, a designer must resort to
detection techniques, which are similar to multiuser detection
methods \cite{verdu}.

The optimal maximum likelihood (ML) detector is described by
\begin{equation}
\hat{\boldsymbol s}_{\rm ML}[i] = \arg \min_{{\boldsymbol s}[i] \in
A} ||{\boldsymbol r}[i] - {\boldsymbol G}{\boldsymbol s}[i]||^2
\end{equation}
where the $KN_U \times 1$ data vector ${\boldsymbol s}[i]$ has the
symbols of all users stacked and the $K N_U \times N_A$ channel
matrix ${\boldsymbol G} = [{\boldsymbol G}_1 \ldots {\boldsymbol
G}_{K}]$ contains the channels of all users concatenated. The ML
detector has a cost that is exponential in the number of data
streams and the modulation order which is too costly for systems
with a large number of antennas. Even though the ML solution can be
alternatively computed using sphere decoder (SD) algorithms
\cite{damen}-\cite{shim} that are very efficient for MIMO systems
with a small number of antennas, the cost of SD algorithms depends
on the noise variance, the number of data streams to be detected and
the signal constellation, resulting in high computational costs for
low SNR values, high-order constellations and a large number of data
streams.

The high computational cost of the ML detector and the SD algorithms
in scenarios with large arrays have motivated the development of
numerous alternative strategies for MIMO detection, which are based
on the computation of receive filters and interference cancellation
strategies. The key advantage of these approaches with receive
filters is that the cost is typically not dependent on the
modulation, the receive filter is computed only once per data packet
and performs detection with the aid of decision thresholds.
Algorithms that can compute the parameters of receive filters with
low cost are of central importance to massive MIMO systems. In what
follows, we will briefly review some relevant suboptimal detectors,
which include linear and decision-driven strategies.

Linear detectors \cite{duel_mimo} include approaches based on the
receive matched filter (RMF), zero forcing (ZF) and minimum
mean-square error (MMSE) designs that are described by
\begin{equation}
\hat{\boldsymbol s}[i] = Q\big( {\boldsymbol W}^H{\boldsymbol r}[i]
\big),
\end{equation}
where the receive filters are
\begin{equation}
{\boldsymbol W}_{\rm RMF} = {\boldsymbol G},~{\rm for ~ the ~ RMF},
\end{equation}
\begin{equation}
{\boldsymbol W}_{\rm MMSE} = {\boldsymbol G}({\boldsymbol
G}^H{\boldsymbol G} + \sigma_s^2/\sigma_n^2 {\boldsymbol I})^{-1},
~{\rm for ~ the~ MMSE ~ design},
\end{equation}
and
\begin{equation}
{\boldsymbol W}_{\rm ZF} = {\boldsymbol G}({\boldsymbol
G}^H{\boldsymbol G})^{-1}, ~{\rm for ~the ~ ZF~ design},
\end{equation}
and $Q(\cdot)$ represents the slicer used for detection.

Decision-driven detection algorithms such as successive interference
cancellation (SIC) approaches used in the {Vertical-Bell
Laboratories Layered Space-Time (VBLAST)} systems
\cite{vblast}-\cite{peng_twc} and decision feedback (DF)
\cite{choi}-\cite{reuter} detectors are techniques that can offer
attractive trade-offs between performance and complexity. Prior work
on SIC and DF schemes has been reported with DF detectors with SIC
(S-DF) \cite{vblast}-\cite{peng_twc} and DF receivers with parallel
interference cancellation (PIC) (P-DF)
\cite{woodward2,delamare_mber,mdfpic}, combinations of these schemes
and mechanisms to mitigate error propagation
\cite{stspadf,reuter,delamare_itic}.

SIC detectors \cite{vblast}-\cite{peng_twc} apply linear receive
filters to the received data followed by subtraction of the
interference and subsequent processing of the remaining users.
Ordering algorithms play an important role as they significantly
affect the performance of SIC receivers. Amongst the existing
criteria for ordering are those based on the channel norm, the SINR,
the SNR and on exhaustive search strategies. The performance of
exhaustive search strategies is the best followed by SINR-based
ordering, SNR-based ordering and channel norm-based ordering,
whereas the computational complexity of an exhaustive search is by
far the highest, followed by SINR-based ordering, SNR-based ordering
and channel norm-based ordering. The data symbol of each user is
detected according to:
\begin{equation}
\hat{s}_k[i] = Q \big( {\boldsymbol w}^H_k{\boldsymbol r}_k[i]
\big),
\end{equation}
where the successively cancelled received data vector that follows a
chosen ordering in the k-th stage is given by
\begin{equation}
{\boldsymbol r}_k[i] = {\boldsymbol r}[i] - \sum_{j=1}^{k-1}
{\boldsymbol g}_j \hat{ s}_{j}[i],
\end{equation}
where ${\boldsymbol g}_j$ corresponds to the columns of the
composite channel matrix ${\boldsymbol G}_j$. After subtracting the
detected symbols from the received signal vector, the remaining
signal vector is processed either by an MMSE or a ZF receive filter
for the data estimation of the remaining users. The computational
complexity of the SIC detector based on either the MMSE or the ZF
criteria is similar and requires a cubic cost in $N_A$ ($O(N_A^3)$)
although the performance of MMSE-based receive filters is superior
to that of ZF-based detectors.

A generalization of SIC techniques, the multi-branch successive
interference cancellation (MB-SIC) algorithm, employs multiple SIC
algorithms in parallel branches. The MB-SIC algorithm relies on
different ordering patterns and produces multiple candidates for
detection, approaching the performance of the ML detector. The
ordering of the first branch is identical to a standard SIC
algorithm and could be based on the channel norm or the SINR,
whereas the remaining branches are ordered by shifted orderings
relative to the first branch. In the $\ell$-th branch, the MB-SIC
detector successively detects the symbols given by the vector
$\boldsymbol{\hat{s}}_\ell[i]=[\hat{s}_{\ell,1}[i],\hat{s}_{\ell,2}[i],\ldots,\hat{s}_{\ell,K}[i]]^T$.
The term $\boldsymbol{\hat{s}}_\ell[i]$ represents the $K \times 1$
ordered estimated symbol vector, which is detected according to the
ordering pattern $\mathbf{T}_\ell,\ell=1,\ldots,S$ for the $\ell$-th
branch. The interference cancellation performed on the received
vector ${\boldsymbol{r}}[i]$ is described by:
\begin{equation}
{\boldsymbol{r}}_{\ell,k}[i]=\boldsymbol{r}[i] -
    \sum_{j=1}^{k-1}\boldsymbol{g}_{\ell,j} \hat{s}_{\ell,j}[i]
\end{equation}
where the transformed channel column $\boldsymbol{g}$ is obtained by
$\boldsymbol{g}_\ell = \boldsymbol{T}_\ell \boldsymbol{g}$, the term
$\boldsymbol{g}'_k$ represents the $k$-th column of the ordered
channel $\boldsymbol{G}'$ and $\hat{s}_{\ell,k}$ denotes the
estimated symbol for each data stream obtained by the MB-SIC
algorithm.

At the end of each branch we can transform ${\hat{\boldsymbol
s}}_\ell[i]$ back to the original order
$\tilde{\boldsymbol{s}}_\ell[i]$ by using $\boldsymbol{T}_\ell$ as
$\tilde{\boldsymbol{s}}_\ell[i]=\boldsymbol{T}_\ell^T\hat{\boldsymbol{{s}}}_\ell[i]$.
At the end of the MB-SIC structure, the algorithm selects the branch
with the minimum Euclidean distance according to
\begin{equation}
\ell_{opt} = \arg \min_{1\leq \ell \leq S} {\boldsymbol C}(\ell)
\end{equation}
where ${\boldsymbol C}(\ell)= || {\boldsymbol r}[i] -
\boldsymbol{T}_\ell {\boldsymbol G} \tilde{\boldsymbol{s}}_\ell[i]
||$ is the Euclidean distance for the $\ell$-th branch. The final
detected symbol vector is
\begin{equation}
\hat{s}_{j}[i] = Q({\boldsymbol w}^H_{\ell_{opt},j} {\boldsymbol
r}_{\ell_{opt},j}[i]), ~~j=1, \ldots, KN_U.
\end{equation}
The MB-SIC algorithm can bring a close-to-optimal performance,
however, the exhaustive search of $S = K!$ branches is not
practical. Therefore, a reduced number of branches $S$ must be
employed. In terms of computational complexity, the MB-SIC algorithm
requires $S$ times the complexity of a standard SIC algorithm.
However, it is possible to implement it using a multi-branch
decision feedback structure \cite{spa,fa} that is equivalent in
performance but which only requires a single matrix inversion as
opposed to $K$ matrix inversions required by the standard SIC
algorithm and $SK$ matrix inversions required by the MB-SIC
algorithm.

DF detectors employ feedforward and feedback matrices that perform
interference cancellation as described by
\begin{equation}
\hat{\boldsymbol s} = Q\big( {\boldsymbol W}^H{\boldsymbol r}[i] -
{\boldsymbol F}^H\hat{\boldsymbol s}_o[i] \big),
\end{equation}
where $\hat{\boldsymbol s}_o$ corresponds to the initial decision
vector that is usually performed by the linear section represented
by ${\boldsymbol W}$ of the DF receiver (e.g., $\hat{\boldsymbol
s}_o = Q( {\boldsymbol W}^H{\boldsymbol r})$) prior to the
application of the feedback section ${\boldsymbol F}$, which may
have a strictly lower triangular structure for performing successive
cancellation or zeros on the main diagonal when performing parallel
cancellation. The receive filters ${\boldsymbol W}$ and
${\boldsymbol F}$ can be computed using various parameter estimation
algorithms which will be discussed in the next section.
Specifically, the receive filters can be based on the RMF, ZF and
MMSE design criteria.

An often criticized aspect of these sub-optimal schemes is that they
typically do not achieve the full receive-diversity order of the ML
algorithm. This led to the investigation of detection strategies
such as lattice-reduction (LR) schemes \cite{windpassinger,gan}, QR
decomposition, M-algorithm (QRD-M) detectors \cite{kim_qrdm},
probabilistic data association (PDA) \cite{jia,syang}, multi-branch
\cite{spa,mbdf} detectors and likelihood ascent search techniques
\cite{vardhan,li}, which can approach the ML performance at an
acceptable cost for moderate to large systems. The development of
cost-effective detection algorithms for massive MIMO systems is a
challenging topic that calls for new approaches and ideas in this
important research area.

\subsection{Iterative Detection and Decoding Techniques}

\begin{figure}[!htb]
\begin{center}
\def\epsfsize#1#2{1\columnwidth}
\epsfbox{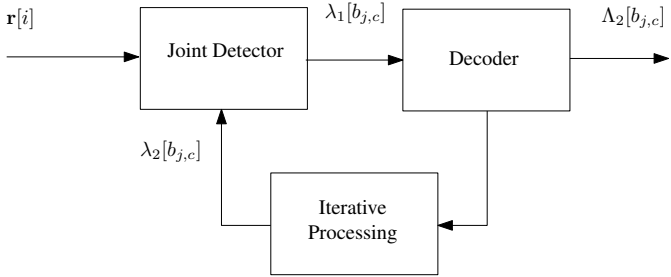} \caption{Block diagram of an IDD scheme.}
\label{idd}
\end{center}
\end{figure}

Iterative detection and decoding (IDD) techniques have received
considerable attention in the last years following the discovery of
Turbo codes \cite{berrou} and the application of the Turbo principle
to interference mitigation
\cite{berrou,douillard,wang,tuchler,hochwald,hou,lee,wu,yuan,choi}.
More recently, work on IDD schemes has been extended to low-density
parity-check codes (LDPC) \cite{hochwald} and \cite{wu} and their
extensions which compete with Turbo codes. The goal of an IDD system
is to combine an efficient soft-input soft-output (SISO) detection
algorithm and a SISO decoding technique as illustrated in Fig.
\ref{idd}. Specifically, the detector produces log-likelihood ratios
(LLRs) associated with the encoded bits and these LLRs serve as
input to the decoder. Then, in the second phase of the
detection/decoding iteration, the decoder generates a posteriori
probabilities (APPs) after a number of (inner) decoding iterations
for encoded bits of each data stream. These APPs are fed to the
detector to help in the next iterations between the detector and the
decoder, which are called outer iterations. The joint process of
detection/decoding is then repeated in an iterative manner until the
maximum number of (inner and outer) iterations is reached. In mobile
cellular networks, a designer can employ convolutional, Turbo or
LDPC codes in IDD schemes for interference mitigation. LDPC codes
exhibit some advantages over Turbo codes that include simpler
decoding and implementation issues. However, LDPC codes often
require a higher number of decoding iterations which translate into
delays or increased complexity. The development of IDD schemes and
decoding algorithms that perform message passing with reduced delays
are of fundamental importance in massive MIMO systems because they
will be able to cope with audio and 3D video which are delay
sensitive.

The massive MIMO systems described at the beginning of this chapter
are considered here with convolutional codes and an iterative
receiver structure consists of the following stages: A
soft-input-soft-output (SISO) detector and a maximum \textit{a
posteriori} (MAP) decoder. Extensions to other channel codes are
straightforward. These stages are separated by interleavers and
deinterleavers. The soft outputs from the detector are used to
estimate LLRs which are interleaved and serve as input to the MAP
decoder for the convolutional code. The MAP decoder \cite{wang}
computes \textit{a posteriori} probabilities (APPs) for each
stream's encoded symbols, which are used to generate soft estimates.
These soft estimates are subsequently used to update the receive
filters of the detector, de-interleaved and fed back through the
feedback filter. The detector computes the \textit{a posteriori}
log-likelihood ratio (LLR) of a symbol ($+1$ or $-1$) for every code
bit of each data stream in a packet with $P$ symbols as given by
\begin{equation}
\begin{split}
\Lambda_1[b_{j,c}[i]] & = {\rm log}
\frac{P[b_{j,c}[i]=+1|{\boldsymbol r}[i]]}{P[b_{j,c}[i]=-1|
{\boldsymbol r}[i]]}, \\  j & =1, \ldots,  K N_U, ~c=1, \ldots, C,
\end{split}
\end{equation}
where $C$ is the number of bits used to map the constellation. Using
Bayes' rule, the above equation can be written as
\begin{equation}
\begin{split}
\Lambda_1[b_{j,c}[i]] & = {\rm log} \frac{P[{\boldsymbol
r}|b_{j,c}[i]=+1]}{P[ {\boldsymbol r}[i]|b_{j,c}[i]=-1]} + {\rm log}
\frac{P[b_{j,c}[i]=+1]}{P[b_{j,c}[i]=-1]}
\\ & = \lambda_1[b_{j,c}[i]] + \lambda_2^p[b_{j,c}[i]], \label{llr}
\end{split}
\end{equation}
where $\lambda_2^p[b_{j,c}[i]] = {\rm log}
\frac{P[b_{j,c}[i]=+1]}{P[b_{j,c}[i]=-1]}$ is the \textit{a priori}
LLR of the code bit $b_{j,c}[i]$, which is computed by the MAP
decoder processing the $j$th data/user stream in the previous
iteration, interleaved and then fed back to the detector. The
superscript $^p$ denotes the quantity obtained in the previous
iteration. Assuming equally likely bits, we have
$\lambda_2^p[b_{j,c}[i]] =0$ in the first iteration for all
streams/users. The quantity $\lambda_1[b_{j,c}[i]] = {\rm log}
\frac{P[{\boldsymbol r}[i]|b_{j,c}[i]=+1]}{P[ {\boldsymbol
r}[i]|b_{j,c}[i]=-1]}$ represents the \textit{extrinsic} information
computed by the SISO detector based on the received data
${\boldsymbol r}[i]$, and the prior information about the code bits
$\lambda_2^p[b_{j,c}[i]],~ j=1,\ldots, KN_U,~ c=1,\ldots, C$ and the
$i$th data symbol. The extrinsic information $\lambda_1[b_{j,c}[i]]$
is obtained from the detector and the prior information provided by
the MAP decoder, which is de-interleaved and fed back into the MAP
decoder of the $j$th data/user stream as the \textit{a priori}
information in the next iteration.

For the MAP decoding, we assume that the interference plus noise at
the output ${\boldsymbol z}_{j}[i]$ of the receive filters is
Gaussian. This assumption has been reported in previous works and
provides an efficient and accurate way of computing the extrinsic
information. Thus, for the $j$th stream/user and the $q$th iteration
the soft output of the detector is
\begin{equation}
z_{j}^{(q)}[i] = V_{j}^{(q)} s_{j}[i] + \xi_{j}^{(q)}[i],
\label{output}
\end{equation}
where $V_{j}^{(q)}[i]$ is a scalar variable equivalent to the
magnitude of the channel corresponding to the $j$th data stream and
$\xi_{j}^{(q)}[i]$ is a Gaussian random variable with variance
$\sigma^2_{\xi_{j}^{(q)}}[i]$. Since we have
\begin{equation}
V_{j}^{(q)} = E\big[ s_{j}^*[i] z_{j}^{(q)[i]} \big]
\end{equation}
and
\begin{equation}
\sigma^2_{\xi_{j}^{(q)}}  = E\big[ | z_{j}^{(q)}[i] - V_{j}^{(q)}[i]
s_{j}[i]|^2 \big],
\end{equation}
the receiver can obtain the estimates ${\hat V}_{j}^{(q)}$ and
${\hat \sigma}^2_{\xi_{j}^{(q)}}$ via corresponding sample averages
{over the received symbols}. These estimates are used to compute the
\textit{a posteriori} probabilities $P[b_{j,c}[i] = \pm 1 |
z_{j,l}^{(q)}[i]]$ which are de-interleaved and used as input to the
MAP decoder. In what follows, it is assumed that the MAP decoder
generates APPs $P[b_{j,c}[i] = \pm 1]$, which are used to compute
the input to the receiver. From (\ref{output}) the extrinsic
information generated by the iterative receiver is given by
\begin{equation}
\begin{split}
\lambda_1[b_{j,c}][i]  & = {\rm log}
\frac{P[z_{j}^{(q)}|b_{j,c}[i]=+1]}{ P[z_{j}^{(q)}|b_{j,c}[i]=-1] }, ~~i=1, \ldots, P, \\
& = \log \frac{\sum\limits_{{\mathbf S} \in {\mathbf S}_c^{+1}} \exp
\Big( -\frac{|z_{j}^{(q)} - V_{j}^{(q)}{\mathbf S}|^2}{ 2
\sigma^2_{\xi_{j}^{(q)}}} \Big)} {\sum\limits_{{\mathbf S} \in
{\mathbf S}_c^{-1}} \exp \Big( -\frac{|z_{j}^{(q)} -
V_{j}^{(q)}{\mathbf S}|^2}{ 2 \sigma^2_{\xi_{j}^{(q)}}} \Big)},
\end{split}
\end{equation}
where ${\mathbf S}_c^{+1}$ and ${\mathbf S}_c^{-1}$ are the sets of
all possible constellations that a symbol can take on such that the
$c$th bit is $1$ and $-1$, respectively. Based on the trellis
structure of the code, the MAP decoder processing the $j$th data
stream computes the \textit{a posteriori} LLR of each coded bit as
described by
\begin{equation}
\begin{split}
\Lambda_2[b_{j,c}[i]]  & = {\rm log} \frac{P[b_{j,c}[i]=+1|
\lambda_1^p[b_{j,c}[i]; {\rm decoding}]}{P[b_{j,c}[i]=-1|
\lambda_1^p[b_{j,c}[i]; {\rm decoding}]} \\ & =
\lambda_2[b_{j,c}[i]] + \lambda_1^p[b_{j,c}[i]], \\ {\rm for} & \;
j=1, \ldots, KN_U, ~c=1, \dots, C.
\end{split}
\end{equation}
The computational burden can be significantly reduced using the
max-log approximation. From the above, it can be seen that the
output of the MAP decoder is the sum of the prior information
$\lambda_1^p[b_{j,c}[i]]$ and the extrinsic information
$\lambda_2[b_{j,c}[i]]$ produced by the MAP decoder. This extrinsic
information is the information about the coded bit $b_{j,c}[i]$
obtained from the selected prior information about the other coded
bits $\lambda_1^p[b_{j,c}[i]], ~ j \neq k$. The MAP decoder also
computes the \textit{a posteriori} LLR of every information bit,
which is used to make a decision on the decoded bit at the last
iteration. After interleaving, the extrinsic information obtained by
the MAP decoder $\lambda_2[b_{j,c}[i]]$ for $j=1, \ldots KN_U$,
$c=1, \dots, C$ is fed back to the detector, as prior information
about the coded bits of all streams in the subsequent iteration. For
the first iteration, $\lambda_1[b_{j,c}[i]]$ and
$\lambda_2[b_{j,c}[i]]$ are statistically independent and as the
iterations are computed they become more correlated and the
improvement due to each iteration is gradually reduced. It is well
known in the field of IDD schemes that there is no performance gain
when using more than $5-8$ iterations.

The choice of channel coding scheme is fundamental for the
performance of iterative joint detection schemes. More sophisticated
schemes than convolutional codes such as Turbo or LDPC codes can be
considered in IDD schemes for the mitigation of multi-beam and other
sources of interference. LDPC codes exhibit some advantages over
Turbo codes that include simpler decoding and implementation issues.
However, LDPC codes often require a higher number of decoding
iterations which translate into delays or increased complexity. The
development of IDD schemes and decoding algorithms that perform
message passing with reduced delays
\cite{wainwright,wymeersch,jingjing} are of great importance in
massive MIMO systems.

\section{Parameter Estimation Techniques}

Amongst the key problems in the uplink of multiuser massive MIMO
systems are the estimation of parameters such as channels gains and
receive filter coefficients of each user as described by the signal
models in Section II. The parameter estimation task usually relies
on pilot (or training) sequences, the structure of the data for
blind estimation and signal processing algorithms. In multiuser
massive MIMO networks, non-orthogonal training sequences are likely
to be used in most application scenarios and the estimation
algorithms must be able to provide the most accurate estimates and
to track the variations due to mobility within a reduced training
period. Standard MIMO linear MMSE and least-squares (LS) channel
estimation algorithms \cite{biguesh} can be used for obtaining CSI.
However, the cost associated with these algorithms is often cubic in
the number of antenna elements at the receiver, i.e., $N_A$ in the
uplink. Moreover, in scenarios with mobility the receiver will need
to employ adaptive algorithms \cite{haykin} which can track the
channel variations. Interestingly, massive MIMO systems may have an
excess of degrees of freedom that translates into a reduced-rank
structure to perform parameter estimation. This is an excellent
opportunity that massive MIMO offers to apply reduced-rank
algorithms \cite{sun}-\cite{jio_mimo} and further develop these
techniques. In this section, we review several parameter estimation
algorithms and discuss several aspects that are specific for massive
MIMO systems such as TDD operation, pilot contamination and the need
for scalable estimation algorithms.

\subsection{TDD operation}

One of the key problems in modern wireless systems is the
acquisition of CSI in a timely way. In time-varying channels, TDD
offers the most suitable alternative to obtain CSI because the
training requirements in a TDD system are independent of the number
of antennas at the base station (or access point)
\cite{rusek,jose,ashikmin} and there is no need for CSI feedback. In
particular, TDD systems rely on reciprocity by which the uplink
channel estimate is used as an estimate of the downlink channel. An
issue in this operation mode is the difference in the transfer
characteristics of the amplifiers and the filters in the two
directions. This can be addressed through measurements and
appropriate calibration \cite{rogalin}. In contrast, in a frequency
division duplexing (FDD) system the training requirements is
proportional to the number of antennas and CSI feedback is
essential. For this reason, massive MIMO systems will most likely
operate in TDD mode and will require further investigation in
calibration methods.

\subsection{Pilot contamination}

The adoption of TDD mode and uplink training in massive MIMO systems
with multiple cells results in a phenomenon called pilot
contamination. In multi-cell scenarios, it is difficult to employ
orthogonal pilot sequences because the duration of the pilot
sequences depends on the number of cells and this duration is
severely limited by the channel coherence time due to mobility.
Therefore, non-orthogonal pilot sequences must be employed and this
affects the CSI employed at the transmitter. Specifically, the
channel estimate is contaminated by a linear combination of channels
of other users that share the same pilot \cite{jose,ashikmin}.
Consequently, the detectors, precoders and resource allocation
algorithms will be highly affected by the contaminated CSI.
Strategies to control or mitigate pilot contamination and its
effects are very important for massive MIMO networks. 

\subsection{Estimation of Channel Parameters}

Let us now consider channel estimation techniques for multiuser
Massive MIMO systems and employ the signal models of Section II. The
channel estimation problem corresponds to solving the following
least-squares (LS) optimization problem:
\begin{equation}
\hat{\boldsymbol G}[i] = \arg \min_{{\boldsymbol G}[i]}
\sum_{l=1}^{i} \lambda^{i-l} ||{\boldsymbol r}[l] - {\boldsymbol
G}[i] {\boldsymbol s}[l]||^2, \label{cest}
\end{equation}
where the $N_A \times K N_U$ matrix ${\boldsymbol G} = [{\boldsymbol
G}_1 \ldots {\boldsymbol G}_K]$ contains the channel parameters of
the $K$ users, the $K N_U \times 1$ vector contains the symbols of
the $K$ users stacked and $\lambda$ is a forgetting factor chosen
between $0$ and $1$. In particular, it is common to use known pilot
symbols ${\boldsymbol s}[i]$ in the beginning of the transmission
for estimation of the channels and the other receive parameters.
This problem can be solved by computing the gradient terms of
(\ref{cest}), equating them to a zero matrix and manipulating the
terms which yields the LS estimate
\begin{equation}
\hat{\boldsymbol G}[i] = {\boldsymbol Q}[i] {\boldsymbol R}^{-1}[i],
\end{equation}
where ${\boldsymbol Q}[i] = \sum_{l=1}^{i} \lambda^{i-l}{\boldsymbol
r}[l] {\boldsymbol s}^H[l]$ is a $N_A \times KNU$ matrix with
estimates of cross-correlations between the pilots and the received
data ${\boldsymbol r}[i]$ and ${\boldsymbol R}[i] = \sum_{l=1}^{i}
\lambda^{i-l} {\boldsymbol s}[i] {\boldsymbol s}^H[i]$ is an
estimate of the auto-correlation matrix of the pilots. When the
channel is static over the duration of the transmission, it is
common to set the forgetting factor $\lambda$ to one. In contrast,
when the channel is time-varying one needs to set $\lambda$ to a
value that corresponds to the coherence time of the channel in order
to track the channel variations.

The LS estimate of the channel can also be computed recursively by
using the matrix inversion lemma \cite{haykin,biguesh}, which yields
the recursive LS (RLS) channel estimation algorithm \cite{fa}
described by
\begin{equation}
{\boldsymbol P}[i] = \lambda^{-1} {\boldsymbol P}[i-1] -
\frac{\lambda^{-2}{\boldsymbol P}[i-1]{\boldsymbol s}[i]
{\boldsymbol s}^{H}[i] {\boldsymbol P}[i-1]}{1+\lambda^{-1}
{\boldsymbol s}^{H}[i] {\boldsymbol P}[i-1]}{\boldsymbol s}[i],
\end{equation}
\begin{equation}
{\boldsymbol T}[i] = \lambda {\boldsymbol T}[i-1] + {\boldsymbol
r}[i]{\boldsymbol s}^H[i],
\end{equation}
\begin{equation}
\hat{\boldsymbol G}[i] = {\boldsymbol T}[i] {\boldsymbol P}[i],
\end{equation}
where the computational complexity of the RLS channel estimation
algorithm is $N_A(KN_U)^2 + 4(KN_U)^2 + 2N_A(KN_U) + 2KN_U + 2$
multiplications and $N_A(KN_U)^2 + 4(KN_U)^2 - KN_U$ additions
\cite{fa}.

An alternative to using LS-based algorithms is to employ least-mean
square (LMS) techniques \cite{sm_ce}, which can reduce the
computational cost. Consider the mean-square error (MSE)-based
optimization problem:
\begin{equation}
\hat{\boldsymbol G}[i] = \arg \min_{{\boldsymbol G}[i]} E
||{\boldsymbol r}[i] - {\boldsymbol G}[i] {\boldsymbol s}[i]||^2],
\label{cest_mse}
\end{equation}
where $E[\cdot]$ stands for expected value. This problem can be
solved by computing the instantaneous gradient terms of
(\ref{cest_mse}), using a gradient descent rule and manipulating the
terms which results in the LMS channel estimation algorithm given by
\begin{equation}
\hat{\boldsymbol G}[i+1] = \hat{\boldsymbol G}[i] + \mu {\boldsymbol
e}[i]{\boldsymbol s}^H[i],
\end{equation}
where the error vector signal is ${\boldsymbol e}[i] = {\boldsymbol
r}[i] - \hat{\boldsymbol G}[i]{\boldsymbol s}[i]$ and the step size
$\mu$ should be chosen between $0$ and $2/tr[{\boldsymbol R}]$
\cite{haykin}. The cost of the LMS channel estimation algorithm in
this scheme is $N_A(KN_U)^2 + N_A(KN_U) + KN_U$ multiplications and
$N_A(KN_U)^2+N_AKN_U +N_A-KN_U$ additions. The LMS approach has a
cost that is one order of magnitude lower than the RLS but the
performance in terms of training speed is worse. The channel
estimates obtained can be used in the ML rule for ML detectors and
SD algorithms, and also to design the receive filters of ZF and MMSE
type detectors outlined in the previous section.

\subsection{Estimation of Receive Filter Parameters}

An alternative to channel estimation techniques is the direct
computation of the receive filters using LS techniques or adaptive
algorithms. In this subsection, we consider the estimation of the
receive filters for multiuser Massive MIMO systems and employ again
the signal models of Section II. The receive filter estimation
problem corresponds to solving the LS optimization problem described
by
\begin{equation}
{\boldsymbol w}_{k,o}[i] = \arg \min_{{\boldsymbol w}_k[i]}
\sum_{l=1}^{i} \lambda^{i-l} |{s}_k[l] - {\boldsymbol w}^H_k[i]
{\boldsymbol r}[l]|^2, \label{fest}
\end{equation}
where the $N_A \times 1$ vector ${\boldsymbol w}_k$ contains the
parameters of the receive filters for the $k$th data stream, the
symbol $s_k[i]$ contains the symbols of the $k$th data stream.
Similarly to channel estimation, it is common to use known pilot
symbols ${s}_k[i]$ in the beginning of the transmission for
estimation of the receiver filters. This problem can be solved by
computing the gradient terms of (\ref{fest}), equating them to a
null vector and manipulating the terms which yields the LS estimate
\begin{equation}
{\boldsymbol w}_{k,o}[i] =  {\boldsymbol R}^{-1}_r[i]{\boldsymbol
p}_k[i],
\end{equation}
where  ${\boldsymbol R}_r[i] = \sum_{l=1}^{i} \lambda^{i-l}
{\boldsymbol r}[i] {\boldsymbol r}^H[i]$ is the auto-correlation
matrix of the received data and ${\boldsymbol p}_k[i] =
\sum_{l=1}^{i} \lambda^{i-l}{\boldsymbol r}[l] {s}^H_k[l]$ is a $N_A
\times 1$ vector with cross-correlations between the pilots and the
received data ${\boldsymbol r}[i]$. When the channel is static over
the duration of the transmission, it is common to set the forgetting
factor $\lambda$ to one. Conversely, when the channel is
time-varying one needs to set $\lambda$ to a value that corresponds
to the coherence time of the channel in order to track the channel
variations. In these situations, a designer can also compute the
parameters recursively, thereby taking advantage of the previously
computed LS estimates and leading to the RLS algorithm \cite{haykin}
given by
\begin{equation}
{\boldsymbol k}[i] = \frac{\lambda^{-1}{\boldsymbol
P}[i-1]{\boldsymbol r}[i]}{1+ \lambda^{-1} {\boldsymbol
r}^H[i]{\boldsymbol P}[i-1]{\boldsymbol r}[i]},
\end{equation}
\begin{equation}
{\boldsymbol P}[i] = \lambda^{-1} {\boldsymbol P}[i-1] -
\lambda^{-1} {\boldsymbol k}[i]{\boldsymbol r}^H[i] {\boldsymbol
P}[i-1],
\end{equation}
\begin{equation}
{\boldsymbol w}_k[i] = {\boldsymbol w}_k[i-1] - {\boldsymbol k}[i]
e_{k,a}^*[i],
\end{equation}
where $e_{k,a}[i] = s_k[i] - {\boldsymbol w}^H_k[i-1]{\boldsymbol
r}[i]$ is the {\it a priori} error signal for the $k$th data stream.
Several other variants of the RLS algorithm could be used to compute
the parameters of the receive filters \cite{sm_tvb}. The
computational cost of this RLS algorithm for all data streams
corresponds to $KN_U (3N^2_A + 4N_A + 1)$ multiplications and $KN_U
(3N_A^2 + 2N_A -1)+2N_AKN_U$ additions

A reduced complexity alternative to the RLS algorithms is to employ
the LMS algorithm to estimate the parameters of the receive filters.
Consider the mean-square error (MSE)-based optimization problem:
\begin{equation}
{\boldsymbol w}_{k,o}[i] = \arg \min_{{\boldsymbol w}_k[i]} E[
|{s}_k[i] - {\boldsymbol w}^H_k[i] {\boldsymbol r}[i]|^2],
\label{fest_mse}
\end{equation}
Similarly to the case of channel estimation, this problem can be
solved by computing the instantaneous gradient terms of
(\ref{fest_mse}), using a gradient descent rule and manipulating the
terms which results in the LMS estimation algorithm given by
\begin{equation}
\hat{\boldsymbol w}_k[i+1] = \hat{\boldsymbol w}_k[i] + \mu {
e}^*_k[i]{\boldsymbol r}[i],
\end{equation}
where the error signal for the $k$th data stream is ${\boldsymbol
e}_k[i] = {s}_k[i] - {\boldsymbol w}^H_k[i]{\boldsymbol r}[i]$ and
the step size $\mu$ should be chosen between $0$ and
$2/tr[{\boldsymbol R}]$ \cite{haykin}. The cost of the LMS
estimation algorithm in this scheme is $KN_U (N_A+1)$
multiplications and $KN_U N_A$ additions.

In parameter estimation problems with a large number of parameters
such as those found in massive MIMO systems, an effective technique
is to employ reduced-rank algorithms which perform dimensionality
reduction followed by parameter estimation with a reduced number of
parameters. Consider the mean-square error (MSE)-based optimization
problem:
\begin{equation}
\big[\bar{\boldsymbol w}_{k,o}[i], {\boldsymbol T}_{D,k,o} [i] \big]
= \arg \min_{\bar{\boldsymbol w}_k[i], {\boldsymbol T}_{D,k}} E[
|{s}_k[i] - \bar{\boldsymbol w}^H_k[i] {\boldsymbol T}_{D,k}^H[i]
{\boldsymbol r}[i]|^2], \label{frrest_mse}
\end{equation}
where ${\boldsymbol T}_{D,k}[i]$ is an $N_A \times D$ matrix that
performs dimensionality reduction and $\bar{\boldsymbol w}_k[i]$ is
a $D \times 1$ parameter vector. Given ${\boldsymbol T}_{D,k}[i]$, a
generic reduced-rank RLS algorithm \cite{jio_mimo} with
$D$-dimensional quantities can be obtained from (39)-(41) by
substituting the $N_A \times 1$ received vector ${\boldsymbol r}[i]$
by the reduced-dimension $D \times 1$ vector $\bar{\boldsymbol
r}[i]={\boldsymbol T}_{D,k}^H[i]{\boldsymbol r}[i]$.


A central design problem is how to compute the dimensionality
reduction matrix ${\boldsymbol T}_{D,k}[i]$ and several techniques
have been considered in the literature, namely:
\begin{itemize}
\item{Principal components (PC): ${\boldsymbol T}_{D,k}[i] = {\boldsymbol
\phi}_{D}[i]$, where ${\boldsymbol \phi}_{D}[i]$ corresponds to a
unitary matrix whose columns are the $D$ eigenvectors corresponding
to the $D$ largest eigenvectors of an estimate of the covariance
matrix $\hat{\boldsymbol R}[i]$.}

\item{Krylov subspace techniques: ${\boldsymbol T}_{D,k}[i] = [ {\boldsymbol
t}_k[i] \hat{\boldsymbol R}[i]{\boldsymbol t}_k[i] \ldots
\hat{\boldsymbol R}^{D-1}[i]{\boldsymbol t}_k[i] $, where
${\boldsymbol t}_k[i] = \frac{{\boldsymbol t}_k[i]}{||{\boldsymbol
p}_k[i]||}$, for $k = 1, 2, \ldots, D$ correspond to the bases of
the Krylov subspace \cite{qian}-\cite{song}. }

\item{Joint iterative optimization methods: ${\boldsymbol T}_{D,k}[i]$
is estimated along with $\bar{\boldsymbol w}_k[i]$ using an
alternating optimization strategy and adaptive algorithms
\cite{delamaresp}-\cite{barc}. }

\end{itemize}

\section{Simulation Results}

In this section, we illustrate some of the techniques outlined in
this article using massive MIMO configurations, namely, a very large
antenna array, an excess of degrees of freedom provided by the array
and a large number of users with multiple antennas. We consider QPSK
modulation, data packets of $1500$ symbols and channels that are
fixed during each data packet and that are modeled by complex
Gaussian random variables with zero mean and variance equal to
unity. For coded systems and iterative detection and decoding, a
non-recursive convolutional code with rate $R=1/2$, constraint
length $3$, generator polynomial $g = [ 7~ 5 ]_{\rm oct}$ and $4$
decoding iterations is adopted. The numerical results are averaged
over $10^6$ runs . For the CAS configuration, we employ $L_k = 0.7$,
$\tau =2$, the distance $d_k$ to the BS is obtained from a uniform
discrete random variable between $0.1$ and $0.95$ , the shadowing
spread is $\sigma_k = 3$ dB and the transmit and receive correlation
coefficients are equal to $\rho = 0.2$. The signal-to-noise ratio
(SNR) in dB per receive antenna is given by $\textrm{SNR} = 10
\log_{10} \frac{KN_U \sigma_{s_r}^2}{R
 C ~\sigma^2}$, where $\sigma_{s_r}^2 = \sigma_s^2 E[|\gamma_k|^2]$ is the variance of the
received symbols, $\sigma^2_n$ is the noise variance, $R<1$ is the
rate of the channel code and $C$ is the number of bits used to
represent the constellation. For the DAS configuration, we use
$L_{k,j}$  taken from a uniform random variable between $0.7$ and
$1$, $\tau =2$, the distance $d_{k,j}$ for each link to an antenna
is obtained from a uniform discrete random variable between $0.1$
and $0.5$ , the shadowing spread is $\sigma_{k,j} = 3$ dB and the
transmit and receive correlation coefficients for the antennas that
are co-located are equal to $\rho = 0.2$. The signal-to-noise ratio
(SNR) in dB per receive antenna for the DAS configuration is given
by $\textrm{SNR} = 10 \log_{10} \frac{KN_U \sigma_{s_r}^2}{R
 C ~\sigma^2}$, where $\sigma_{s_r}^2 = \sigma_s^2 E[|\gamma_{k,j}|^2]$ is the variance of the
received symbols.

\begin{figure}[!htb]
\begin{center}
\def\epsfsize#1#2{1\columnwidth}
\epsfbox{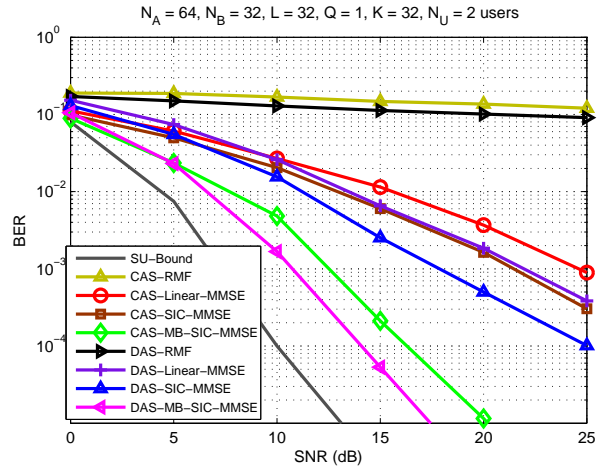} \vspace*{-0.5em} \caption{BER performance against
SNR of detection algorithms in a scenario with $N_A=64$, $N_B = 32$,
$L = 32$, $Q = 1$, $K=32$ users and $N_U=2$ antenna elements.}
\label{ber}
\end{center}
\end{figure}

In the first example, we compare the BER performance against the SNR
of several detection algorithms, namely, the RMF with multiple users
and with a single user denoted as single user bound, the linear MMSE
detector \cite{verdu}, the SIC-MMSE detector using a successive
interference cancellation \cite{rontogiannis} and the multi-branch
SIC-MMSE (MB-SIC-MMSE) detector \cite{spa,fa,mbdf}. We assume
perfect channel state information and synchronization. In
particular, a scenario with $N_A=64$ antenna elements at the
receiver, $K=32$ users and $N_U=2$ antenna elements at the user
devices is considered, which corresponds to a scenario without an
excess of degrees of freedom with $N_A \approx K N_U$. The results
shown in Fig. \ref{ber} indicate that the RMF with a single user has
the best performance, followed by the MB-SIC-MMSE, the SIC-MMSE, the
linear MMSE and the RMF detectors. Unlike previous works
\cite{rusek} that advocate the use of the RMF, it is clear that the
BER performance loss experienced by the RMF should be avoided and
more advanced receivers should be considered. However, the cost of
linear and SIC receivers is dictated by the matrix inversion of $N_A
\times N_A$ matrices which must be reduced for large systems.
Moreover, it is clear that a DAS configuration is able to offer a
superior BER performance due to a reduction of the average distance
from the users to the receive antennas and a reduced correlation
amongst the set of $N_a$ receive antennas, resulting in improved
links.

\begin{figure}[!htb]
\begin{center}
\def\epsfsize#1#2{1\columnwidth}
\epsfbox{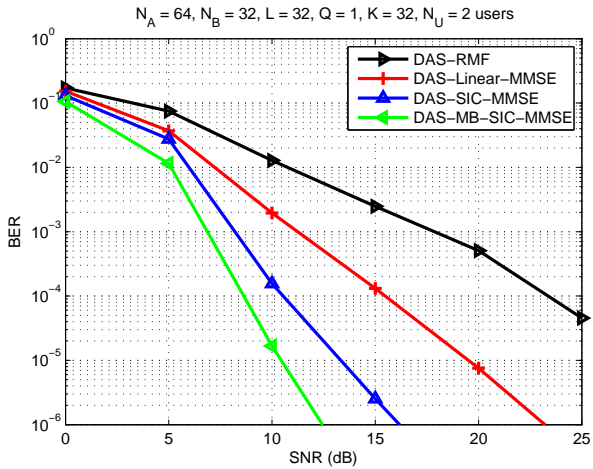} \vspace*{-0.5em} \caption{Coded BER performance
against SNR of detection algorithms with DAS in a scenario with
$N_A=64$, $N_B = 32$, $L = 32$, $Q = 1$, $K=32$ users, $N_U=2$
antenna elements and $4$ iterations.} \label{cber}
\end{center}
\end{figure}

In the second example, we consider the coded BER performance against
the SNR of several detection algorithms with a DAS configuration
using perfect channel state information, as illustrated in Fig.
\ref{cber}. The results show that the BER is much lower than that
obtained for an uncoded systems as indicated in Fig. \ref{ber}.
Specifically, the MB-SIC-MMSE algorithm obtains the best performance
followed by the SIC-MMSE, the linear MMSE and the RMF techniques.
Techniques like the MB-SIC-MMSE and SIC-MMSE are more promising for
systems with a large number of antennas and users as they can
operate with lower SNR values and are therefore more energy
efficient. Interestingly, the RMF can offer a BER performance that
is acceptable when operating with a high SNR that is not energy
efficient and has the advantage that it does not require a matrix
inversion. If a designer chooses stronger channel codes like Turbo
and LDPC techniques, this choice might allow the operation of the
system at lower SNR values.

\begin{figure}[!htb]
\begin{center}
\def\epsfsize#1#2{1\columnwidth}
\epsfbox{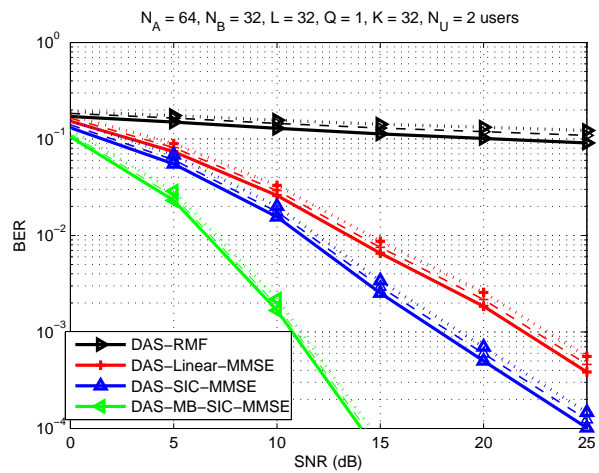} \vspace*{-0.5em} \caption{BER performance against
SNR of detection algorithms in a scenario with channel estimation,
$N_A=64$, $N_B = 32$, $L = 32$, $Q = 1$, $K=32$ users and $N_U=2$
antenna elements. Parameters: $\lambda = 0.999$ and $\mu = 0.05$.
The solid lines correspond to perfect channel state information, the
dashed lines correspond to channel estimation with the RLS algorithm
and the dotted lines correspond to channel estimation with the LMS
algorithm.} \label{ber_ce}
\end{center}
\end{figure}

In the third example, we assess the estimation algorithms when
applied to the analyzed detectors. In particular, we compare the BER
performance against the SNR of several detection algorithms with a
DAS configuration using perfect channel state information and
estimated channels with the RLS and the LMS algorithms. The channels
are estimated with $250$ pilot symbols which are sent at the
beginning of packets with $1500$ symbols. The results shown in Fig.
\ref{ber_ce} indicate that the performance loss caused by the use of
the estimated channels is not significant as it remains within
$1$-$2$ dB for the same BER performance. The main problems of the
use of the standard RLS and LMS is that they require a reasonably
large number of pilot symbols to obtain accurate estimates of the
channels, resulting in reduced transmission efficiency.

\begin{figure}[!htb]
\begin{center}
\def\epsfsize#1#2{1\columnwidth}
\epsfbox{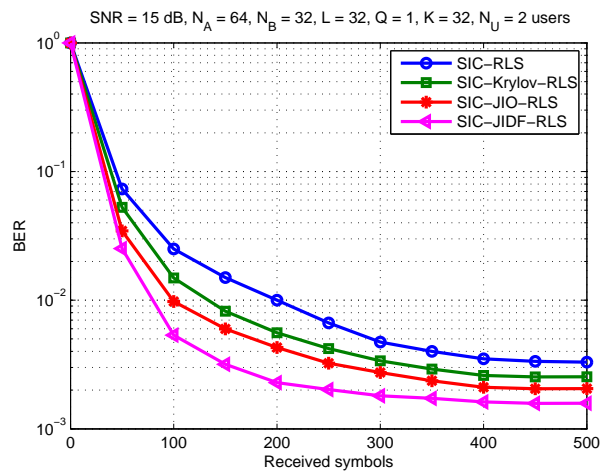} \vspace*{-0.5em} \caption{BER performance against
the number of received symbols for SIC receivers with a DAS
architecture operating at SNR = 15 dB in a scenario with the
estimation of the receive filters, $N_A=64$, $N_B = 32$, $L = 32$,
$Q = 1$, $K=32$ users and $N_U=2$ antenna elements. Parameters:
$\lambda = 0.999$, $D=5$ for all reduced-rank methods, and
interpolators with $I=3$ parameters and $12$ branches for the JIDF
scheme.} \label{berxsymbols}
\end{center}
\end{figure}

In the fourth example, we evaluate the more sophisticated
reduced-rank estimation algorithms to reduce the number of pilot
symbols for the training of the receiver filters. In particular, we
compare the BER performance against the number of received symbols
for a SIC type receiver using a DAS configuration and the standard
RLS \cite{haykin}, the Krylov-RLS \cite{honig} and JIO-RLS
\cite{jio_mimo} and the JIDF-RLS \cite{jidf} algorithms. We provide
the algorithms pilots for the adjustment of the receive filters and
assess the BER convergence performance. The results shown in Fig.
\ref{berxsymbols} illustrate that the performance of the
reduced-rank algorithms is significantly better than the standard
RLS algorithm, indicating that the use of reduced-rank algorithms
can reduce the need for pilot symbols. Specifically, the best
performance is obtained by the JIDF-RLS algorithm, followed by the
JIO-RLS, the Krylov-RLS and the standard RLS techniques. In
particular, the reduced-rank algorithms can obtain a performance
comparable to the standard RLS algorithm with a fraction of the
number of pilot symbols required by the RLS algorithm. It should be
remarked that for larger training periods the standard RLS algorithm
will converge to the MMSE bound and the reduced-rank algorithms
might converge to the MMSE bound or to higher MSE values depending
on the structure of the covariance matrix ${\boldsymbol R}$ and the
choice of the rank $D$.

\section{Future Trends and Emerging Topics}

In this section, we discuss some future signal detection and
estimation trends in the area of massive MIMO systems and point out
some topics that might attract the interest of researchers. The
topics are structured as:

\begin{itemize}

\item{Signal detection:\\

$\rightarrow$ Cost-effective detection algorithms: Techiques to
perform dimensionality reduction \cite{goldstein}-\cite{barc} for
detection problems will play an important role in massive MIMO
devices. By reducing the number of effective processing elements,
detection algorithms could be applied. In addition, the development
of schemes based on RMF with non-linear interference cancellation
capabilities might be a promising option that can close the
complexity gap
between RMF and more costly detectors.\\

$\rightarrow$ Decoding strategies with low delay: The development of
decoding strategies for DAS configurations with reduced delay will
play a key role in applications such as audio and video streaming
because of their delay sensitivity. Therefore, we novel message
passing algorithms with smarter strategies to exchange information
should be investigated along with their application to IDD schemes \cite{wainwright,wymeersch,jingjing}. \\

$\rightarrow$ Mitigation of impairments: The identification of
impairments originated in the RF chains of massive MIMO systems,
delays caused by DAS schemes will need mitigation by smart signal
processing algorithms. For example, I/Q imbalance might be dealt
with using widely-linear signal processing algorithms
\cite{chevalier,song,song2}.\\

$\rightarrow$ Detection techniques for multicell scenarios: The
development of detection algorithms for scenarios with multiple and
small cells requires approaches which minimize the need for channel
state information from adjacent cells and the decoding delay
\cite{dai}-\cite{rmpdid}.\\ }

\item{Parameter estimation:\\

$\rightarrow$ Blind algorithms: The development of blind estimation
algorithms for the channel and receive filter parameters is
important for mitigating the problem of pilot contamination \cite{honig1995blind}-\cite{ccmavf}. \\

$\rightarrow$ Reduced-rank and sparsity-aware algorithms: the
development of reduced-rank and sparsity-aware algorithms that
exploit the mathematical structure of massive MIMO channels is an
important topic for the future along with features that lend
themselves to implementation \cite{qian}-\cite{barc}. }

\end{itemize}

\section{Concluding Remarks}

This chapter has presented signal detection and estimation
techniques for multiuser massive MIMO systems. We consider the
application to cellular networks with massive MIMO along with CAS
and DAS configurations. Recent signal detection algorithms have been
discussed and their use with iteration detection and decoding
schemes has been considered. Parameter estimation algorithms have
also been reviewed and studied in several scenarios of interest.
Numerical results have illustrated some of the discussions on signal
detection and estimation techniques along with future trends in the
field.

\end{document}